\documentstyle[aps,prb,graphics,epsf,multicol]{revtex}

\begin{document}

\title{Two-dimensional Dilute Ising Models: Defect
Lines and the Universality of the Critical Exponent $\nu$}

\author{Ferenc Szalma$^{1}$ and Ferenc Igl\'oi$^{2,1}$}

\address{
$^1$ Institute for Theoretical Physics,
Szeged University, H-6720 Szeged, Hungary\\
$^2$ Research Institute for Solid State Physics and Optics, 
H-1525 Budapest, P.O.Box 49, Hungary\\
}

\date{October 25, 1998}

\maketitle

\begin{abstract}
We consider two-dimensional Ising models with randomly distributed
ferromagnetic bonds and study the local critical behavior at defect
lines by extensive Monte Carlo simulations. Both for {\it ladder} and
{\it chain} type defects, non-universal critical behavior is observed:
the critical exponent of the defect magnetization is found 
to be a continuous function of the
strength of the defect coupling. Analyzing corresponding stability
conditions, we obtain new evidence that the critical exponent $\nu$ of
the bulk correlation length of the random Ising model does not depend 
on dilution, i.e. $\nu=1$.

\end{abstract}

\vspace{5mm}



\newcommand{\bc}{\begin{center}}
\newcommand{\ec}{\end{center}}
\newcommand{\be}{\begin{equation}}
\newcommand{\ee}{\end{equation}}
\newcommand{\beqn}{\begin{eqnarray}}
\newcommand{\eeqn}{\end{eqnarray}}

\begin{multicols}{2}
\narrowtext

The presence of quenched randomness may drastically change the
critical properties of magnetic systems. For disorder which
is coupled to the energy density, i.e. in particular for random bond 
and random site dilution, the relevance-irrelevance of the perturbation at a
second-order phase transition point is given by the well known Harris
criterion\cite{harris}. If the specific heat exponent of the pure
system is positive, $\alpha>0$, a new random fixed point is expected
to control the critical properties of the dilute model. The marginal
situation in the Harris criterion, $\alpha=0$, is represented by the
two-dimensional (2d) Ising model, in which case detailed studies,
both (field-)theoretical\cite{dot1,sha1} and numerical\cite{sel1,RAJ,SAQS},
have been performed to
clarify the critical properties of the dilute model. By now, according
to general view, the dilution is considered as a marginally irrelevant
perturbation, thus the critical singularities in the dilute Ising model are
characterized by the power laws of the perfect model modified by
logarithmic corrections. There is, however, another view which
interprets numerical data as giving evidence for dilution dependent critical
exponents\cite{kuhn}.

In this paper we try to decide between the conflicting views in an
indirect way by studying the local critical behavior at a defect
line\cite{ipt} in the dilute model. A defect line, which could be located at
grain boundaries in real systems, represents a marginal perturbation
in the 2d perfect Ising model. According to Bariev's exact
solution\cite{bariev} the critical exponent $\beta_d$, defined via the 
temperature dependence of the defect or local magnetization $m_d$,
\be
m_d \sim t^{\beta_d},~~~t=(T_c-T)/T_c \to 0^+\;,
\label{betad}
\ee
is a continuous function of the strength of the defect coupling $J_d$.
\end{multicols}
\widetext
\begin{figure}[t]
{\centering \resizebox*{.7\columnwidth}{!}{\includegraphics{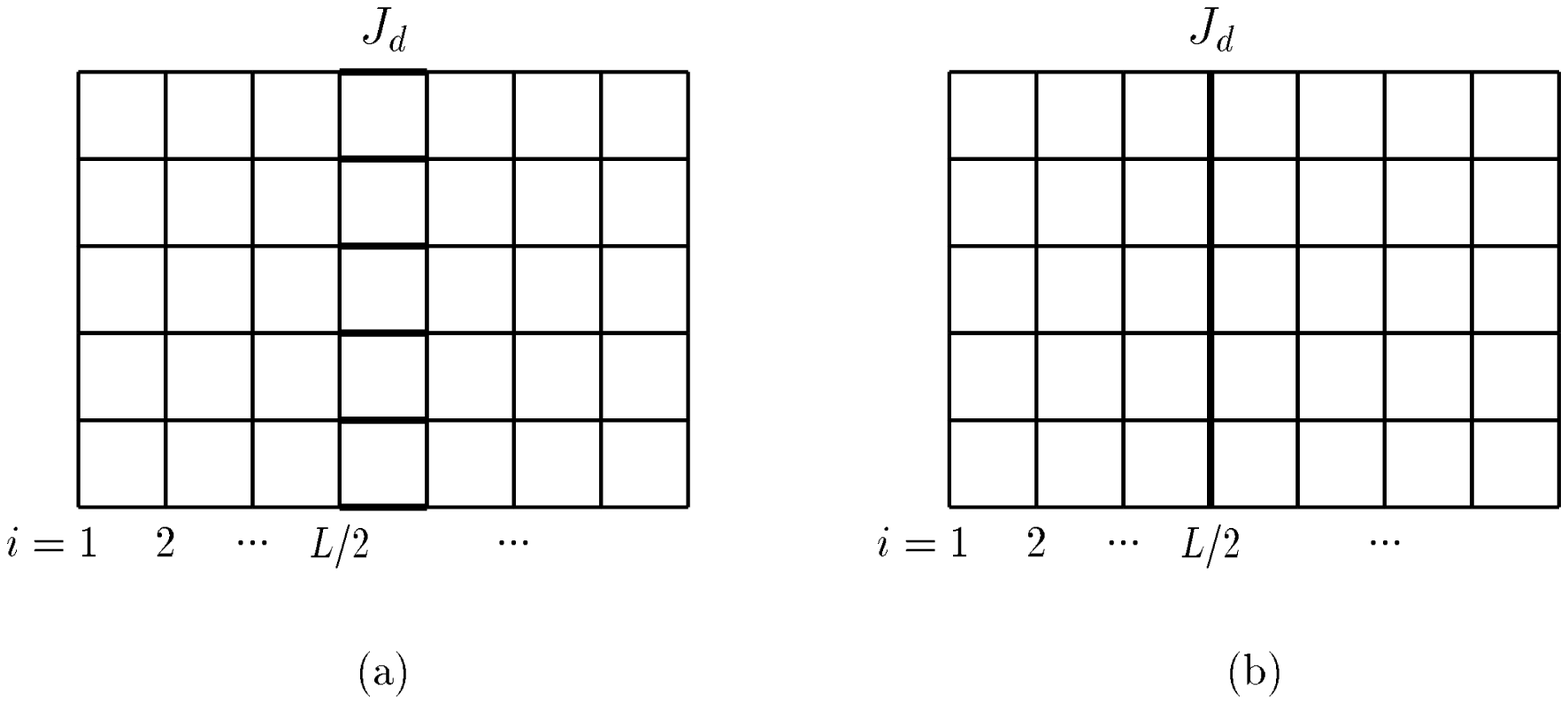}} \par}
\caption{Ladder (a) and chain (b) type defects.}
\label{fig1}
\end{figure}
\begin{multicols}{2}
\narrowtext
For a chain defect, see Fig. 1b, one gets
\be
\beta_d={2 \over \pi^2} \arctan^2 \kappa_c,~~~~\kappa_c=
\exp\left[-2(J_d-J)/T_c\right]\;,
\label{chain}
\ee
whereas for a ladder defect, see Fig. 1a, $\beta_d$ is given by
\be
\beta_d={2 \over \pi^2} \arctan^2 \kappa_l,~~~~\kappa_l=
{{\tanh}(J/T_c) \over {\tanh}(J_d/T_c)}\;,
\label{ladder}
\ee
where $J$ is the coupling in the isotropic Ising model. We note that
the above formulae could be generalized to non-isotropic models,
as well\cite{bariev}. As shown recently by Pleimling and Selke\cite{edge},
the edge magnetization of three-dimensional Ising magnets at the surface
transition has a 
similar non-universal critical behavior, which, indeed, can be related to the
local critical behavior at a defect line in the two-dimensional Ising
model.

The exact results on the local critical behavior of the 2d Ising
model in eqs. (\ref{chain}) and (\ref{ladder}) are in complete agreement
with a stability analysis of the fixed point of the homogeneous system 
in the presence of a defect line\cite{burkh,ipt}.
Under a small perturbation this fixed point is unstable, if the
critical exponent of the bulk correlation length, $\nu$, is $\nu \le 1$.
Furthermore, a ladder defect with small local couplings behaves like 
two weakly 
coupled surfaces, and ordinary surface critical behavior will result,
provided the corresponding surface fixed point remains stable against
a weak coupling between the surfaces\cite{burkh,ipt}. 
This stability condition can
be expressed in terms of the surface susceptibility exponent of the
homogeneous model 
as $\gamma_{1,1} <0$\cite{ipt}. Applying hyperscaling, one obtains for 
$d=2$, $\gamma_{1,1}=\nu-2 \beta_1$, where $\beta_1$ is the critical 
exponent of the surface magnetization).
As one may easily check the corresponding two marginality conditions, 
for ladder defects,
\be
\nu=1~~\text{ and}~~~\gamma_{1,1}=0\;,
\label{marginal}
\ee
are both satisfied for the 2d Ising model; the marginality is manifested
by the defect coupling dependent critical exponent in
eq. (\ref{ladder}).

In the following, we are going to utilize the above observations and study
the local critical behavior at defect lines in the {\it dilute} Ising
model. We consider strongly diluted systems, so that the bulk critical
region is clearly controlled by the random fixed point,
and insert the line defects as local perturbation. Then the
relevance-irrelevance criterion for the local critical behavior is expected
to have the same form as described above, with the
exponents, $\nu$ and $\gamma_{1,1}$, referring now to the dilute model. 
Determining the local magnetization exponent $\beta_d$, at the defect, 
one may imagine two scenarios:
i) $\beta_d$ showing a continuous variation with the defect coupling $J_d$,
or ii) $\beta_d$ staying constant in, at least, some extended range
of $J_d$. In the first case, there would be  evidence that the
marginality conditions, see eq. (\ref{marginal}), remain valid for 
the dilute model.
Otherwise, one might infer that the critical exponents $\nu$ and
$\gamma_{1,1}$ for the pure and dilute models are different.

In what follows we consider a random-bond nearest neighbor Ising model on a
square lattice
where the random ferromagnetic couplings, $J_1$ and $J_2$, occur with
equal probability. That model is self-dual\cite{fisch}, and the
self-duality point
\be
{\tanh}(J_1/T_c)=\exp(-2J_2/T_c)\;,
\label{selfdual}
\ee
corresponds to the critical point, if there is one phase transition in
the system. Indeed, this assumption is strongly supported by  numerical 
calculations\cite{sel1}. In this dilute model,
ladder and chain defects are then introduced, where the defect couplings 
$J_d$ are uniform and ferromagnetic.

To calculate the local critical properties we did extensive
Monte Carlo (MC) simulations using Wolff's cluster flip algorithm\cite{wolf}. 
We have considered square lattices with $L$ columns and $L$ rows; 
$J_d$ couples neighboring spins in the center column for chain defects, 
whereas for the ladder defect $J_d$ connects spins between the two center
columns. Typically we took $L=256$, applying full periodic boundary
conditions and generating about $10^4$ clusters per
realization. The results are then averaged over hundred realizations.
The statistical errors during a MC run in a given sample turned out
to be significantly smaller than those arising from the ensemble
averaging.
We mention that similar parameters were used in the previous
study on the surface critical behavior of the dilute Ising
model\cite{ssli}, which corresponds to the case with a ladder defect,
where $J_d$ vanishes.

In the MC simulations we calculated the average magnetization per
column, $m(i)=\langle | \sum s_{ij} | \rangle /L$, where the sum runs
over $j=1,2,\dots,L$. The defect magnetization is then given by
$m_d=m(L/2)$. The simulations were performed at three values of the
dilution parameter $r=J_1/J_2=1$, $1/4$ and $1/10$ and for several
values of the defect coupling in the region $0 \le J_d/J_2 \le 4$.

The magnetization profile $m(i)$ displays at the defect either a  maximum or a minimum depending on the strength of the defect couplings, 
as illustrated in Fig. 2 for ladder defects.
Far from the defect, there
is a plateau in the profile with the height signaling the bulk magnetization,
$m_b$. The size of the defect region, $l_d$, where the magnetization differs
substantially from its bulk value, is related to the bulk correlation length
of the system.

In the thermodynamic limit, $L \to \infty$, as the critical temperature
$T_c$, see eq. (\ref{selfdual}), is approached, the magnetization 
profile $m(i)$ goes
to zero as a power-law $m(i) \sim t^{\beta(i)}$, where
$\beta(L/2)=\beta_d$ and $\beta(i)=\beta$ for $|L/2-i|>l_d$, where $\beta$ 
is the usual bulk critical exponent. 
To estimate the values of these critical exponents from simulational data, 
one may define  temperature dependent effective exponents
\be
\beta(i)_{eff}=\mathrm{d} \ln[m(i)]/\mathrm{d} \ln[t]\;,
\label{betaeff}
\ee
which are approximated by using  data at discrete temperatures, 
say, $t+\Delta t/2$
and $t-\Delta t/2$. In the limit of sufficiently small $\Delta t$ and $t$, 
the effective exponents approach the true critical exponents, presuming 
that the system is large enough so that finite-size effects play no role.
\begin{figure}

\epsfxsize=\columnwidth\epsfbox{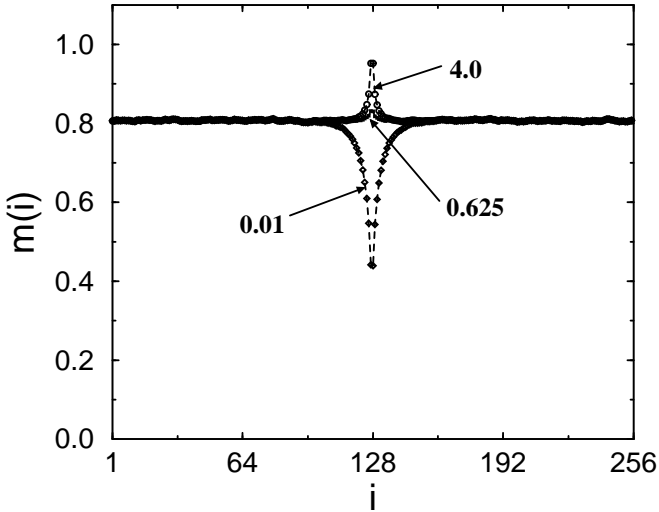}
\caption{Magnetization profiles in the dilute Ising model with a ladder
defect at different couplings $J_d/J_2=$4.0, 0.625 and 0.01 at the
dilution $r=J_1/J_2=1/4$ and reduced critical temperature $t=0.06$. 
MC systems of size $256 \times 256$ were simulated.
\label{fig2}}
\end{figure}

\noindent
To avoid finite-size effects, $L$ should be much larger than the 
correlation lengths in the bulk and
at the defect line. Actually we approached the
critical point by calculating $\beta(i)_{eff}$ for
$t=0.15,~0.13,~0.11,~0.09$ and $0.07$, with $\Delta t =0.02$, and then
did a linear extrapolation to $t=0$. The error caused by 
the extrapolation seems to be rather small.
Further technical details can be found in Ref. 14.

Before presenting our findings about the defect line problem in the
dilute model, we will first consider the perfect model with random defect
couplings. The aim of this part of our investigation is to clarify, whether 
random defects could lead to varying local exponents.
The two random, ferromagnetic couplings in the defect line, $J$ and $J_d$, 
are assumed to occur with equal
probability, where $J$ is also the coupling in the rest of the system.  
Results about the local magnetization exponent $\beta_d$ for various values
of the ratio $J_d/J$ are shown in Fig. 3, both for chain and ladder defects. 
The error bars in Fig. 3 take into account the sample averaging and
the extrapolation. Only a few typical error bars are shown.

For $J_d/J=1$, the critical exponent of the perfect model, $\beta=1/8$,
is reproduced quite accurately. For other values of that ratio, one observes a 
non-universal critical behavior with $\beta_d$ varying continuously 
with $J_d/J$, in accordance with the above marginality conditions.- It 
may be interesting to note the touching of the curves
for the ladder and the chain defects at $J_d=J$  .

\begin{figure}
\epsfxsize=\columnwidth\epsfbox{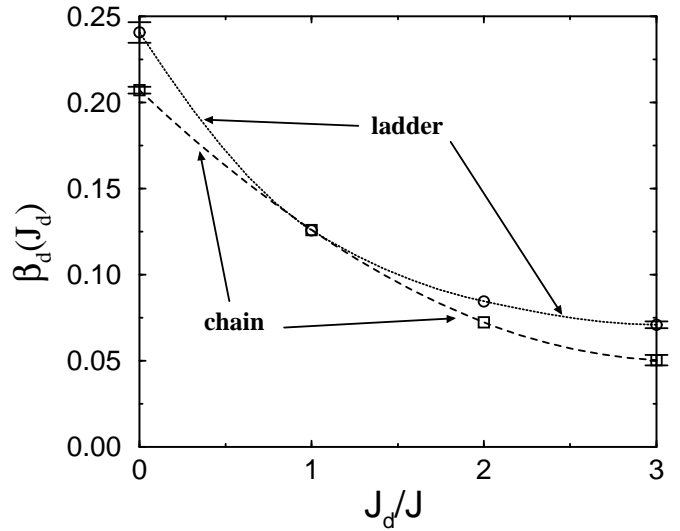}
\caption{Local magnetization exponent, $\beta_d$, of the pure Ising model 
with randomly
distributed couplings in the chain and ladder type defects,
see text. The lines are guides to the eye.}  
\label{fig3}
\end{figure}

In Fig. 4, results about the critical exponent of the local magnetization
$\beta_d$ for the dilute model, with random couplings $J_1$ and $J_2$, and 
the defect line, with uniform coupling $J_d$, are depicted. 

\begin{figure}
\epsfxsize=\columnwidth\epsfbox{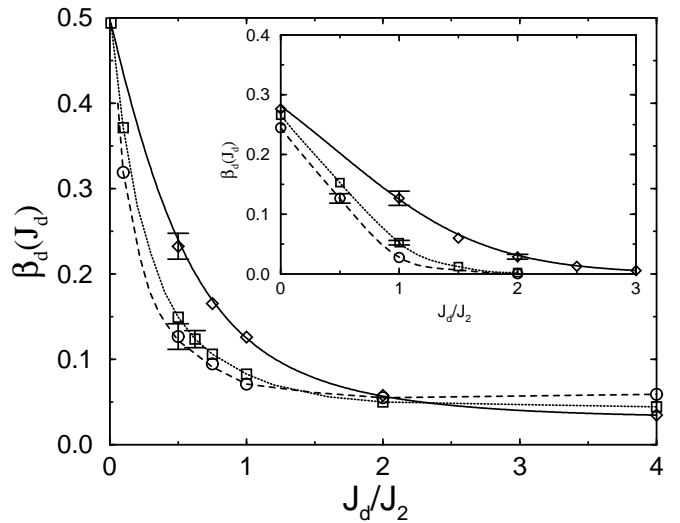}
\caption{Local magnetization exponent, $\beta_d$, of the random Ising model at
different dilutions ($r=J_1/J_2=1$ diamond, $r=1/4$ square, $r=1/10$ circle)
for ladder defects. Findings on the chain defect are shown in the inset.
The exact results for the pure system with $r=1$ are denoted by full line;
the broken and dotted lines are guides to the eye.}  
\label{fig4}
\end{figure}

\noindent
In the
case $r=J_1/J_2=1$, our data are in good agreement with the exact results,
see eqs. (\ref{chain}) and (\ref{ladder}) for chain and ladder defects. 
In the dilute case, $J_1 \neq J_2$, $\beta_d$ is seen to vary continuously 
with the strength of the defect coupling $J_d$. 
For a fixed value of $J_d$,
the defect energy-density increases, relative to the average bulk value,
for decreasing value of $r=J_1/J_2$. Therefore, generally, there is an
increasing local order at the defect, which is connected to a decreasing value of the defect exponent, $\beta_d$. 
This argument, however, seems to be not valid for the
ladder defect with $J_d \gg J_2$. In this limit one has effectively a chain
defect with random couplings $J_1+J_2$, with probability $1/2$, as well as 
$2J_1$ and $2J_2$, each with probability $1/4$. Then, shown in
Fig. 4, $\beta_d(J_d)$ is increasing with increasing dilution.- For $J_d=0$, 
one recovers\cite{ssli,diehl} the surface critical exponent 
$\beta_d=\beta_1=1/2$.

Another limiting situation is obtained for a chain defect with zero
defect bond, $J_d=0$. Then the problem is equivalent to a ladder defect 
with three random couplings, which depend on $J_1$ and $J_2$. 
As seen in the inset of Fig. 4,
such random couplings could also lead to a non-universal behavior.
This observation is in agreement with our
findings on the perfect model in Fig. 3.

To summarize, we considered uniform ladder and chain defects in
two-dimensional dilute Ising models and determined the critical exponent 
of the defect magnetization. The exponent was found to be a continuous 
function of the defect coupling.
Assuming that the first stability criterion mentioned above holds 
for the dilute 
case as well, one gets for the critical exponent $\nu$  of the bulk 
random Ising model the borderline value $\nu=1$. Accordingly, 
one could rule out $\nu>1$, as 
had been suggested before\cite{kuhn} in the context of dilution dependent bulk 
critical exponents. 

In conclusion, we suggest that the non-universal critical behavior is
related to the borderline values of the critical exponents of the 
bulk dilute model, as given in eq. (\ref{marginal}). 
Consequently, one obtains $\nu=1$ and 
$\gamma_{1,1}=0$ (implying $\beta_1=1/2$, in agreement with Ref. 14), both for 
the perfect and dilute two dimensional Ising models .

\acknowledgements

This work has been supported by the
Hungarian National Research Fund under grants No OTKA TO23642 and
OTKA F/7/026004 and by the Ministery of Education under grant
No FKFP 0765/1997. Useful discussions with W. Selke and L. Turban are 
gratefully
acknowledged. F. Sz. thanks the Institut f\"ur Theoretische Physik, Technische
Hochschule Aachen, where part of this work has been completed, for kind
hospitality, and the DAAD for the scholarship enabling his visit there.


\end{multicols}

\end{document}